# On the Performance of Co-existence between Public eMBB and Non-public URLLC Networks

Yanpeng Yang[1], Kimmo Hiltunen[2], and Fedor Chernogorov[2]

{yanpeng.yang, kimmo.hiltunen, fedor.chernogorov}@ericsson.com

Ericsson Research, [1]Sweden, [2]Finland

*Abstract*—To ensure the high level of automation required in today's industrial applications, next-generation wireless networks must enable real-time control and automation of dynamic processes with the requirements of extreme low-latency and ultra-reliable communications. In this paper, we provide a performance assessment for the co-existence of a public enhanced mobile broadband (eMBB) and a local non-public factory (URLLC) network and evaluate the network conditions under which the stringent latency and reliability requirements of factory automation applications are met. The evaluations consider both an unsynchronized and a synchronized time division duplexing (TDD) deployment between the networks, as well as scenarios both with and without any macro eMBB traffic located inside the factory. The results show that an unsynchronized deployment is possible if the isolation between the networks is sufficiently high, either as a result of a separation distance, wall loss or the use of separate frequencies for the networks. A synchronized deployment will avoid the cross-link interference, but it will not resolve the problems related to the closed access and the near-far interference. If the factory contains eMBB traffic served by the overlaid macro cells, the performance of both networks will suffer due to a high level of cross-link and near-far interference. The problems related to the cross-link interference can be resolved by synchronizing the networks, while the level of the near-far interference can be reduced by allowing the eMBB users to be connected to base stations located inside the factory. Finally, if an unsynchronized deployment is desired, the factory should be deployed on an isolated frequency.

*Keywords*—URLLC, co-existence, smart manufacturing, non-public network, spectrum

## I. Introduction

Ultra-reliable low latency communication (URLLC) is expected to support unprecedented levels of high reliability and low latency communications, e.g. industrial automation and control, real-time operation of the smart electrical power grid, and remote control of real-time operations [1]. In order to secure that the 5G New Radio (NR) supports the wide field of URLLC services, the Third Generation Partnership Project (3GPP) has outlined the general URLLC reliability requirement for one transmission of a 32-byte packet with a 99.999% success probability (block error rate (BLER) of $10^{-5}$) and a user plane latency of 1ms [2][3].

One of the envisioned URLLC use cases is factory automation where closed-loop control applications run periodic cycles with short cycle times while demanding extremely low error rate. In addition, it is likely that the factory is located inside the coverage area of a macro network, as shown in Fig. 1. From the co-existence analysis point of view, there are two spectrum options that are of interest: co-channel deployment and adjacent channel deployment. For the co-channel overlay, the networks are operated by the same mobile network operator (MNO), and the same spectrum is reused in both. In case of an adjacent channel deployment, the factory network has been deployed on a separate frequency compared to the macro enhanced mobile broadband (eMBB) network, owned either by the macro MNO, a neighboring MNO, or a local license holder.

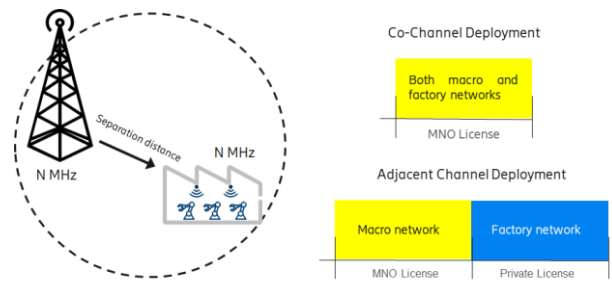

Fig. 1. Assumed co-existence scenario and spectrum allocation.

In order to better understand the feasibility of this kind of co-existence scenario, the impact of the inter-network interference on both the URLLC and the eMBB network performance has to be evaluated. The evaluations should consider different kinds of radio network deployment options to find out which of them would be feasible for each scenario. Looking at the literature, numerous studies assuming different types of co-existence scenarios can be found, where a macro network has to co-exist with another radio network or system. For example, in [4] and [5], interference mitigation and joint power allocation schemes are proposed to enable co-existence between radar and communication systems. The authors in [6] and [7] provide spectrum sharing methods for co-existence between 5G systems and fixed service in the 28GHz band. It is shown that the separation distance between the two systems depends on the antenna pointing directions and the system bandwidths. Another popular topic is the co-existence between long term evolution (LTE) and Wi-Fi in unlicensed band, which can be supported by medium access control (MAC) protocol designs and decoding methods, as presented in [8][9] and the references therein. The co-existence between macro and small cells operating in time-division duplex (TDD) has been studied in [10] and [11], where resource allocation and interference cancellation schemes are investigated to facilitate the heterogeneous deployment. However, none of the listed studies consider scenarios where the victim network is assumed to be offering a URLLC service with stringent latency and reliability requirements. The authors in [12] evaluate the performance of a co-existence scenario between a public 5G NR eMBB macro network and a non-public 5G NR URLLC factory network. The evaluation considers both the impact of unsynchronized TDD (uTDD) and synchronized TDD (sTDD), as well as the impact of the separation distance between the factory and the closest macro site on the network performance. However, the study does not consider the special scenario where eMBB users are located inside the factory. Furthermore, no potential solutions are provided to improve the performance of the co-existence scenario.

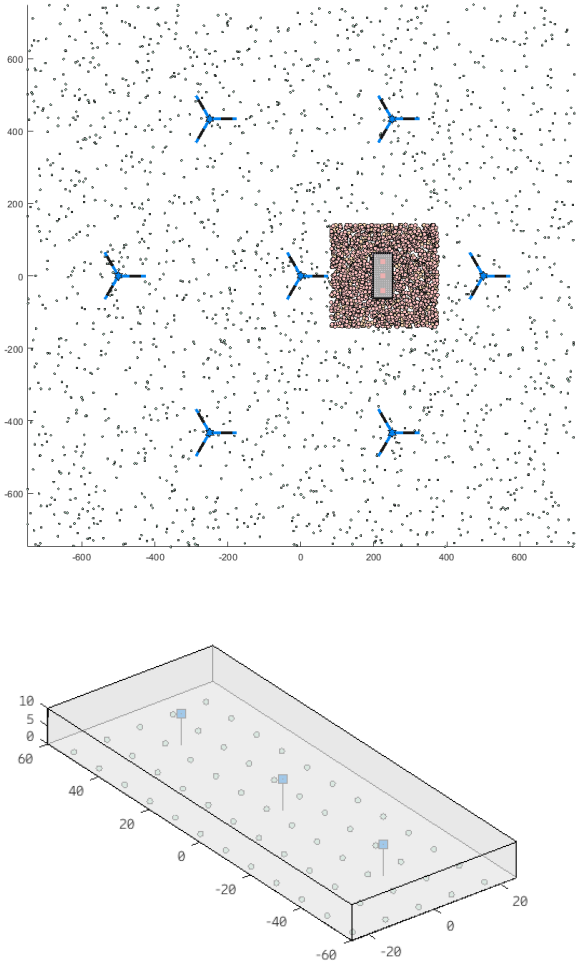

Fig. 2. Network layout of the co-existence scenario and the layout of the factory network.

Motivated by the initial results and the identified research gaps, we substantially extend [12] to provide a more complete picture of the performance of the assumed co-existence scenario. First of all, we evaluate the service availability of the URLLC network and the average user throughput of the eMBB network when the factory contains eMBB traffic served by the overlaid macro cells. Secondly, we propose and evaluate several potential solutions to improve the URLLC performance with and without eMBB traffic inside the factory.

The rest of this paper is organized as follows. In Section II, we describe the system model. Simulation methodology and parameters are introduced in Section III. Simulation results and analysis are presented in Section III. Finally, conclusions are drawn in Section IV.

## II. SYSTEM MODEL

### A. Network layout

Our study is conducted in an area of 1500 m by 1500 m which consists of a two-tier overlaid macro network and a factory network, as shown in Fig. 2. In the macro network, seven tri-sector sites with beamformed antennas are deployed. The sites are 25 m high and have an inter-site distance of 500 m. The factory is located close to the macro cell border, at 200 m distance from the closest macro site. The factory layout is based on the 3GPP factory model with the size of 120x50x10 $m^3$ [3]. By default, three ceiling-mounted omni-directional factory base stations are deployed as illustrated in Fig. 2. The eMBB UEs are randomly distributed in outdoor locations over the evaluated system area, and in some specific cases also inside the factory. The factory is surrounded by an eMBB traffic hotspot ('impact area') to ensure that the factory network can cope with the worst-case scenarios. Due to the eMBB traffic hotspot, the macro base stations close to the factory will operate with a higher level of resource utilization and will therefore generate more interference towards the factory.

### B. Propagation models

We assume the 3GPP Urban Macro (UMa) propagation model for the links between the macro base stations and the outdoor eMBB users, and the 3GPP Indoor Hotspot (InH) model for the links between the factory base stations and the URLLC users [13]. Furthermore, the path losses between the macro base stations and the UEs or base stations inside the factory are calculated as a combination of the UMa propagation model, wall penetration loss and an indoor loss. Finally, the path loss between the factory base stations or UEs and the outdoor eMBB UEs is modeled as a combination of the 3GPP Urban Micro (UMi) propagation model [13], wall penetration loss and an indoor loss. The wall penetration loss consists of two parts: a constant value for a perpendicular penetration and an additional loss depending on the grazing angle [14]. In this study, we assume that the wall penetration loss for perpendicular penetration is equal to 14 dB by default, corresponding to an 18 cm thick slab of concrete [15]. Finally, the indoor loss is expressed as $D \cdot d_{in}$ where $D$ is 0.5 dB/m as in [13] and $d_{in}$ is the travelled indoor distance.

### C. Duplex models

Both networks are assumed to be operating on NR mid-band (4 GHz) and the performance of two different TDD co-existence scenarios is evaluated as illustrated in Fig. 3:

- Unsynchronized TDD: The macro network follows a slot-based DDDU TDD pattern while the factory network follows a sub-slot-based DUDU TDD pattern.

- Synchronized TDD: Both networks follow a sub-slot-based DUDU TDD pattern.

The slot borders are assumed to be aligned for both TDD configurations. It is worth noting that the DDDU pattern is often regulated for eMBB networks on a country level and therefore the DUDU pattern is a highly unlikely option. However, the downside of assuming the eMBB-optimized DDDU pattern for both networks is that it is not able to satisfy the most stringent latency requirements (e.g., the 1 ms requirement assumed in this study) and may face uplink capacity problems even for the more relaxed URLLC services. Hence, in this paper, the URLLC network is always assumed to follow the sub-slot-based DUDU pattern.

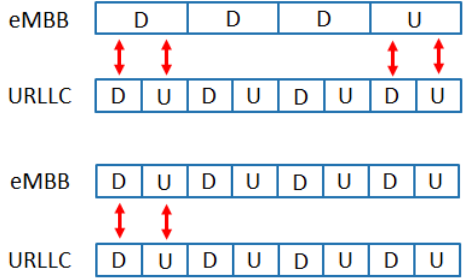

Fig. 3. Assumed TDD patterns for uTDD (top) and sTDD (bottom). The red arrows are indicating the different inter-network interference scenarios.

TABLE I. MAIN SIMULATION PARAMETERS

| Parameter | eMBB network | URLLC network |
|---|---|---|
| Frequency [GHz] | 4 | 4 |
| Bandwidth [MHz] | 50 | 50 |
| TDD pattern | DDDU & DUDU | DUDU |
| Total offered traffic within the system area | 350 Mbps (DL) 117 Mbps (UL) | 40 Mbps (DL) 40 Mbps (UL) |
| Traffic ratio DL:UL | 3:1 | 1:1 |
| Sectors per site | 3 | 1 |
| BS Transmit Power [dBm] | 50 | 30 |
| UE Transmit Power [dBm] | 23 | 23 |
| BS noise figure [dB] | 5 | 5 |
| UE noise figure [dB] | 9 | 9 |
| Max BS Antenna element gain [dBi] | 8 | 2 |
| BS Antenna Array V x H x (Vs x Hs x Ps) | 8x8x(1x1x2) | Omni-directional |
| UE Antenna | Isotropic (0 dBi) | Isotropic (0 dBi) |
| Beamforming scheme | Long term wideband eigen beamforming | No beamforming |
| Uplink power control | SNR based: $\alpha = 0.8$, target SNR = 10 dB | SNR based: $\alpha = 0.8$, target SNR = 10 dB |
| ACIR [dB] | BS-to-UE = 32.7  BS-to-BS = 38.8 | UE-to-BS = 29.6  UE-to-UE = 28.2 |

## III. SIMULATION METHODOLOGY AND PARAMETERS

Table I summarizes the simulation parameters assumed in this study. The offered eMBB downlink traffic volume is assumed to be equal to 150 Mbps within the impact area, and 200 Mbps elsewhere. Assuming the 3:1 traffic split, the corresponding uplink traffic volumes are equal to 50 Mbps and 67 Mbps, respectively.

For the 5G NR URLLC network, a subcarrier spacing of 30 kHz and packet size of 32 Bytes are assumed. A transmission time interval (TTI) length of 143 μs is considered with 4 OFDM symbols per TTI. Moreover, we consider QPSK, 16 QAM, and 64 QAM for the available modulation and coding schemes of the URLLC network with the corresponding (1/20, 1/10, 1/5, 1/3), (1/3, 1/2, 2/3), and (2/3, 3/4) code rates, respectively. During the simulations the offered URLLC traffic is fixed at 40 Mbps for both the downlink and the uplink which is the maximum traffic volume that can be served when the isolation between the networks is set to infinity.

The URLLC users are assumed to be successfully served if they can fulfill the reliability requirement of 99.999% within a one-way latency bound of 1 ms. We select service availability as the URLLC performance metric, which is defined as the percentage of UEs fulfilling the assumed reliability and latency requirements [3]. The desired service availability is assumed to be equal to 100%, which means that the URLLC service requirements have to be fulfilled for each location within the factory floor.

Finally, to evaluate the adjacent channel deployment, the adjacent channel interference ratio (ACIR) values for the different inter-network interference scenarios have been calculated from the adjacent channel leakage ratio (ACLR) and the adjacent channel selectivity (ACS) requirements listed in [16] and [17]. In case of sTDD, only the downlink-to-downlink (BS-to-UE) and the uplink-to-uplink (UE-to-BS) ACIR are needed. However, in case of uTDD, two additional ACIR values, the downlink-to-uplink (BS-to-BS) and the uplink-to-downlink (UE-to-UE) ACIR, are needed as well.

## IV. SIMULATION RESULTS AND ANALYSIS

The assumed co-existence scenario between the macro and the factory network becomes very challenging, if some active eMBB users served by the overlaid public macro cells are located inside the factory. This is due to a few main differences compared to the evaluations in [12]. First and most importantly, the wall penetration loss is no longer helping to mitigate the inter-network interference from the eMBB users towards both the factory base stations and users. In fact, the situation becomes even worse if the wall loss is increased, because as a result the transmission powers will be increased for the eMBB users located inside the factory. Secondly, the downlink beams from the macro base stations will point towards the factory to serve the traffic inside the factory, increasing the level of interference. At the same time, the increased path losses will contribute to an increased average resource utilization within the macro cells, resulting further in a higher level of inter-network interference. Finally, the eMBB and URLLC UEs get closer to each other which increases the level of the cross-link interference as well. From the eMBB network point of view, the users entering the factory are expected to experience greatly reduced throughputs. This is mostly due to the impact of the wall loss, which reduces the received signal power levels and at the same time increases the level of the inter-network interference caused by the factory base stations and UEs. In order to study the impact of moving part of the eMBB traffic into the factory, we set two levels of factory eMBB traffic volume, 1 Mbps and 10 Mbps, to represent a low and a moderate level of the factory eMBB traffic. For comparison, we evaluate also the baseline scenario where no eMBB traffic is located inside the factory. Finally, it should be highlighted that the eMBB throughput results consider only the UEs located inside the factory except for the baseline scenario where the average is calculated for the users located within the 'impact area'. Hence, the focus is on the eMBB users experiencing the highest level of the inter-network interference from the factory network.

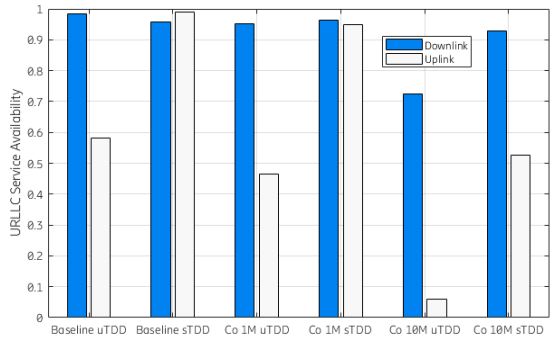

(a) URLLC service availability

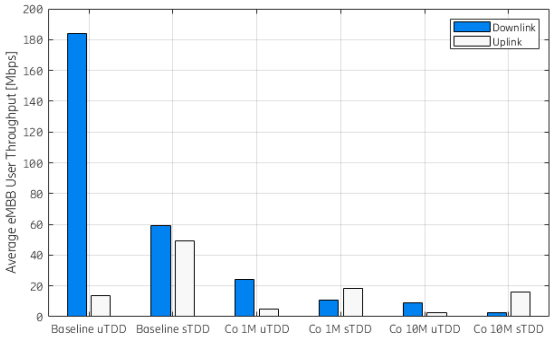

(b) Average eMBB user throughput

Fig. 4. URLLC service availability and the average eMBB user throughput, assuming a co-channel deployment and different levels of eMBB traffic (1 Mbps or 10 Mbps) inside the factory.

*A. Co-channel deployment*

The results for the co-channel deployment are presented in Fig. 4. For the baseline scenario without any eMBB traffic inside the factory, the obtained results show that in case of uTDD, the desired 100% URLLC service availability cannot be reached due to the high level of inter-network interference caused by the high-power macro base stations. Uplink suffers more than downlink because the SNR-based uplink power control is limiting the received signal powers, leading to low signal-to-interference-plus-noise-ratio (SINR) values and a high average resource utilization. In case of sTDD, the level of the uplink inter-network interference becomes very low since it originates from the eMBB UEs with reduced utilization due to an increased amount of allocated time domain resources. However, at the same time the URLLC downlink performance becomes slightly worse compared to the unsynchronized deployment because the eMBB downlink gets less time domain resources resulting in a higher resource utilization and thus, a higher level of inter-network interference towards the factory.

As already discussed, a co-channel deployment between the eMBB and the URLLC network is not expected to be feasible due to severe near-far problems when eMBB UEs enter the factory. The results in Fig. 4(a) demonstrate how the URLLC uplink service availability is dramatically reduced especially at a higher level of the factory eMBB traffic. The performance is worse with uTDD due to the interference from the macro base stations towards the factory base stations while in case of sTDD the uplink is affected only by the interference caused by the eMBB UEs located both inside and outside the factory. Meanwhile, the URLLC downlink performance is affected by the interference from the macro base stations, and in case of uTDD also by the interference from the close-by eMBB UEs with high transmit powers, which makes the service availability lower than in case of sTDD.

Fig. 4(b) indicates that the average eMBB user throughputs within the co-channel deployment are unacceptable for almost all cases due to the high level of inter-network interference and the reduced signal powers from the serving macro base stations. In more detail, the eMBB downlink is affected by the interference from the factory URLLC base stations, and in case of uTDD also by the interference from the URLLC UEs. Similarly, the eMBB uplink is affected by the interference from the URLLC UEs and in case of uTDD also by the interference from the factory base stations.

*B. Adjacent channel deployment*

The situation becomes slightly less challenging for the adjacent channel deployment, as demonstrated by the results in Fig. 5. For the baseline scenario, both the URLLC downlink and uplink reach 100% service availability without any additional mechanisms to mitigate interference. With a low level of eMBB traffic inside the factory, URLLC downlink can still reach 100% service availability with both uTDD and sTDD. However, with a higher level of factory eMBB traffic, sTDD can reach 100% service availability, but uTDD cannot. In case of URLLC uplink, the desired 100% service availability is not reached even with the low level of factory eMBB traffic.

The results in Fig. 5(b) demonstrate that the performance of the eMBB UEs inside the factory is affected for both uTDD and sTDD. Compared to the co-channel deployment, the ACIR reduces the level of the inter-network interference, but it does not affect the impact of wall loss on the received signal power. As can be noticed by comparing Fig. 4(b) and Fig. 5(b), the eMBB downlink performance benefits a lot from the adjacent channel deployment while the uplink performance remains still at a low level. Such phenomenon suggests that the downlink performance of the eMBB users located inside the factory is affected more by the interference while the uplink is mostly limited by the wall loss.

*C. Potential solutions*

Due to the completely different service requirements, the desirable situation would be to be able to avoid any tight coordination or synchronization on the TDD pattern level between the networks. Unfortunately, that kind of scenario would also be the most challenging from the co-existence point of view. Given the results presented in this paper, the URLLC service availability is guaranteed only in an adjacent channel deployment without any eMBB traffic located inside the factory. In order to improve the URLLC performance for the other co-existence scenarios, the level and/or the impact of the inter-network interference has to be reduced.

To start with, a synchronized TDD deployment between the networks will remove the cross-link interference, in particular the interference between the base stations, but it will not resolve the problems related to the closed access and the near-far interference. Furthermore, if the eMBB-optimized TDD pattern is applied for the URLLC network, services with the most stringent latency requirements cannot be supported. In addition, even the URLLC services with more relaxed latency requirements could face problems related to the maximum uplink capacity.

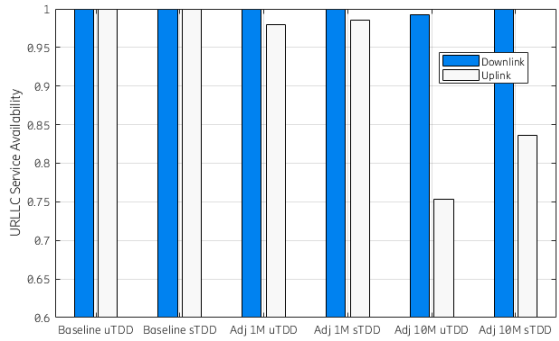

(a) URLLC service availability

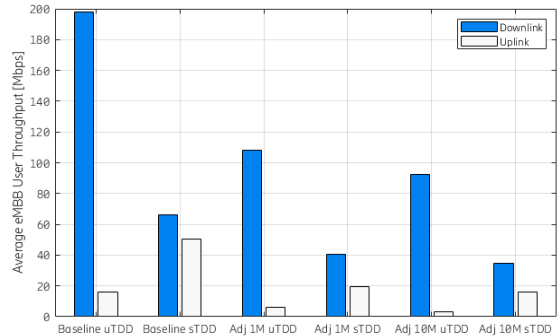

(b) Average eMBB user throughput

Fig. 5. URLLC service availability and the average eMBB user throughput, assuming an adjacent deployment and different levels of eMBB traffic inside the factory.

An efficient way to reduce the near-far interference, and to improve the eMBB user performance, is to allow the eMBB users to connect to the factory network, i.e., by having a shared RAN type of solution within the factory. A possible alternative in case of the adjacent channel deployment would be to deploy a separate public indoor network within the factory. However, in that case a new co-existence scenario is created, where the macro network has to co-exist with both the co-channel public factory network and the adjacent channel non-public factory network. Furthermore, there could be challenges related to the co-existence between the public factory network and the adjacent channel non-public factory network.

The level of the inter-network interference can be reduced by increasing the level of the isolation between the networks for example by increasing the separation distance [12] and the wall loss, by applying a guard band, or by deploying the factory network on an isolated frequency. However, if the factory contains eMBB traffic, an increased isolation with the help of an increased wall loss may make the near-far problems even worse due to the increased transmission powers for the eMBB users located inside the factory. As an example, the impact of wall loss on the URLLC service availability is evaluated in Fig. 6, assuming a co-channel uTDD co-existence scenario without any eMBB traffic inside the factory. As can be noticed, both the downlink and the uplink service availability reach the desired 100% level, when the wall loss is increased from 14 dB to 35 dB, corresponding to a 47 cm thick slab of concrete [15]. Hence, the required wall loss becomes so large that some other wall material (e.g., metal) in addition to concrete should be used.

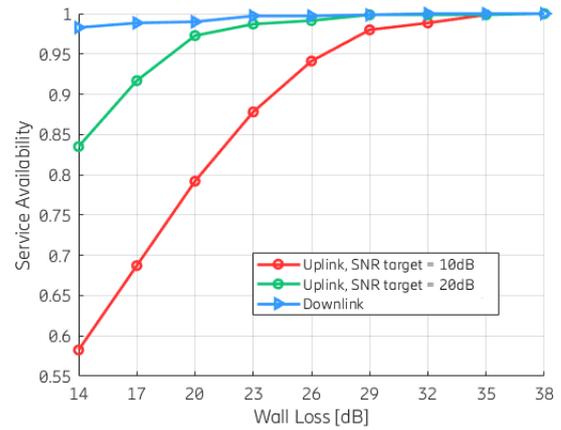

Fig. 6. The impact of wall loss and the increased uplink transmission power on the URLLC service availability, assuming uTDD and no eMBB traffic inside the factory.

The impact of the inter-network interference on the URLLC performance can be reduced by making sure that the factory network is dominating over the macro network within the areas where the URLLC service is desired. This can be achieved for example by increasing the transmission powers of the URLLC base station and the UEs, or by densifying the URLLC network. Furthermore, the use of directional or beamformed base station antennas can help to improve the performance. An example of the impact of increasing the URLLC UE transmission powers is provided in Fig. 6, where the uplink service availability is evaluated as a function of the wall loss, when the SNR target for the power control is increased from 10 dB to 20 dB. As demonstrated by the results, the required wall loss resulting in the desired 100% service availability drops from 35 dB to 29 dB (39 cm thick slab of concrete [15]). In general, this kind of uplink desensitization helps as long as the URLLC UEs will not be transmitting on the maximum power. Furthermore, it may not be able to fully compensate for the strong uplink interference peaks caused by the macro eMBB users located inside the factory. For example, it was not able to resolve the uplink performance problems for the adjacent channel deployment with a low level of macro eMBB traffic inside the factory (see Fig. 5).

The impact of URLLC network densification is evaluated in Fig. 7, where the number of factory base stations is increased from 3 to 12. As can be noticed, the downlink service availability is marginally improved as a result of the network densification, but the desired level of 100% cannot be guaranteed. The situation is even worse for the uplink, where the service availability is reduced as a result of the network densification. In general, network densification is found to be beneficial for noise-limited deployments to secure a sufficient coverage throughout the factory floor. However, in case of interference-limited deployments the network densification will usually lead into reduced SINR values, due to the fact that the level of the inter-cell interference increases faster than the received signal from the serving cell. Even though the amount of available radio resources is increased within the factory, that is not able to compensate for the negative impact of the reduced SINRs, keeping also in mind that the URLLC system performance is typically limited by the users with the worst SINRs. The reason why the situation is even worse for uplink is that due to the power control, the level of the received signal

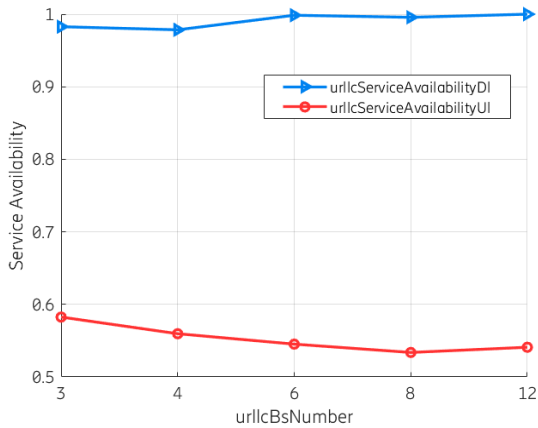

Fig. 7. Impact of network densification on the URLLC service availability, assuming uTDD and no eMBB traffic inside the factory

power is kept roughly constant, while the level of the inter-cell interference is increasing.

## V. CONCLUSIONS AND FUTURE WORK

As a summary for this study, an unsynchronized deployment between the public macro network and the non-public URLLC factory network is very difficult, unless the isolation between the networks is sufficiently large either as a result of a separation distance, wall loss or a frequency separation. A synchronized deployment avoids the cross-link interference between the networks, but may result in capacity problems within the network that has to follow the non-optimized TDD pattern. Furthermore, if the URLLC network is aligned with the eMBB-optimized TDD pattern, services with the most stringent latency requirements cannot be supported. The impact of the inter-network interference on the URLLC uplink performance can be reduced also by increasing the UE transmission power levels within the factory. However, the situation cannot be improved by densifying the URLLC network, because the URLLC performance becomes limited by the inter-cell interference instead of the inter-network interference.

If the factory contains eMBB traffic served by the macro cells, there will be severe problems related both to the cross-link interference and to the near-far interference. Based on the results shown in this paper, an adjacent channel deployment is not sufficient to resolve all the co-existence problems. On top of that, the networks have to be synchronized, and the eMBB users have to be able to connect to base stations located inside the factory. If an unsynchronized deployment is desired, the solution is to deploy the factory network on an isolated frequency.

There are still a few open questions that should be investigated further. To start with, the performance could be re-evaluated assuming more relaxed URLLC requirements, when for example the impact of applying the eMBB-optimized TDD pattern to the URLLC network could be evaluated. Furthermore, a more uplink-heavy URLLC service could be assumed as well, related to e.g., a remote control use case, where a high-quality video stream is transmitted in uplink, while only some low bit rate control information is transmitted in the downlink. The feasibility of the option to deploy a public factory network to avoid the near-far problems should be evaluated in more detail as well. Finally, the impact of network densification on the URLLC performance is a topic that should be investigated further, for example when it comes to the impact of the URLLC service requirements, and the type of the base station antennas or the factory environment.

ACKNOWLEDGMENT

This work has been performed in the framework of the H2020 project 5G-SMART co-funded by the EU. The authors would like to acknowledge the contributions of their colleagues. This information reflects the consortium's view, but the consortium is not liable for any use that may be made of any of the information contained therein.